\documentclass[12pt]{article}
\pdfoutput=1

\usepackage{fullpage}
\usepackage{amsmath,amssymb,hyperref,feynmp-auto}

\newcommand{\ud}{\text{d}}

\begin{document}

\title{Gaugino masses from misaligned supersymmetry breaking and R-symmetry breaking spurions}
\author{Yunhao Fu\textsuperscript{a,\dag},
        Tianjun Li\textsuperscript{b,c,\ddag},
        Longjie Ran\textsuperscript{a,\S},
        Zheng Sun\textsuperscript{a,*}\\
        \textsuperscript{a}%
        \normalsize\textit{College of Physics, Sichuan University,}\\
        \normalsize\textit{29 Wangjiang Road, Chengdu 610064, P.\ R.\ China}\\
        \textsuperscript{b}%
        \normalsize\textit{CAS Key Laboratory of Theoretical Physics,}\\
        \normalsize\textit{Institute of Theoretical Physics, Chinese Academy of Sciences,}\\
        \normalsize\textit{55 Zhongguancun East Street, Beijing 100190, P.\ R.\ China}\\
        \textsuperscript{c}%
        \normalsize\textit{School of Physical Sciences, University of Chinese Academy of Sciences,}\\
        \normalsize\textit{19(A) Yuquan Road, Beijing 100049, P.\ R.\ China}\\
        \normalsize\textit{E-mail:}
        \parbox[t]{26em}{
        \textsuperscript{\dag}\texttt{2021222020008@stu.scu.edu.cn,}
        \textsuperscript{\ddag}\texttt{tli@itp.cas.cn,}\\
        \textsuperscript{\S}\texttt{2019222020004@stu.scu.edu.cn,}
        \textsuperscript{*}\texttt{sun\_ctp@scu.edu.cn}}
       }
\date{}
\maketitle

\begin{abstract}
In gauge mediation models with multiple spurion fields breaking SUSY and the R-symmetry separately, we show that it is possible to generate gaugino masses at one loop if the R-charge arrangement satisfies a certain condition.  The resulting gaugino masses are calculated and suppressed by some power of the messenger mass scale.  We present two simple examples to demonstrate this possibility, and discuss possible phenomenology implications.
\end{abstract}

\section{Introduction}

Gauge mediation~\cite{Martin:1997ns, Giudice:1998bp, Meade:2008wd, Kitano:2010fa} provides a mechanism to mediate spontaneous supersymmetry (SUSY) breaking effects from the SUSY breaking hidden sector to the visible supersymmetric standard model (SSM) sector, giving soft terms for sparticle masses and coupling constants which may be tested in the LHC or other future experiments.  Although recent experimental results have excluded many SUSY models~\cite{ParticleDataGroup:2022pth}, there are still some unexplored corners of the parameter space where models compatible with experiments can be constructed.

In gauge mediation models, SUSY breaking fields in the hidden sector~\cite{Intriligator:2007cp} are coupled to messenger fields which are nontrivial representations of the standard model (SM) gauge symmetries.  The hidden sector is effectively described by a Wess-Zumino model which possesses F-term spontaneous SUSY breaking~\cite{Wess:1973kz, Wess:1974jb}.  A chiral superfield $X$, named the SUSY breaking spurion, obtains a vacuum expectation value (VEV) for its auxiliary field component after spontaneous SUSY breaking.  To build a natural SUSY breaking model in the hidden sector, it is necessary to introduce an R-symmetry which restricts the possible terms appearing in the superpotential~\cite{Nelson:1993nf}.  The R-charge assignment of fields determines the generic form of the superpotential which has R-charge $2$, and SUSY or SUSY breaking vacua can be obtained by a proper R-charge assignment satisfying certain conditions~\cite{Sun:2011fq, Kang:2012fn, Sun:2019bnd, Amariti:2020lvx, Li:2020wdk, Sun:2021svm, Li:2021ydn, Brister:2021xca, Brister:2022rrz, Brister:2022vsz, Sun:2022xdl}.  SUSY breaking effects are then transferred to the visible sector through the coupling of messenger fields.  In gauge mediation models, the R-symmetry also need to be spontaneously broken to generate Majorana gaugino masses which do not respect the R-symmetry.  Effectively, The R-symmetry is spontaneously broken by the VEV of the scalar component of a chiral superfield $Y$ with a nonzero R-charge, which is named the R-symmetry breaking spurion.  In a simple model of the hidden sector, the SUSY breaking and R-symmetry breaking spurions are usually considered to be aligned, and $X$ is identified with $Y$.  The scalar component of $X$, or the SUSY breaking pseudomodulus $\phi_X$~\cite{Ray:2006wk, Sun:2008nh, Sun:2011aq}, obtains a VEV at loop level by including quantum corrections~\cite{Shih:2007av, Curtin:2012yu}, or through the inclusion of D-terms~\cite{Azeyanagi:2012pc, Vaknin:2014fxa}.  Thus $X$ has the VEV
\begin{equation} \label{eq:1-01}
\langle X \rangle = \langle \phi_X \rangle
                    + \theta^2 \langle F_X \rangle \ .
\end{equation}
Sparticle masses are generated through the gauge mediation mechanism.  Among them, the gauginos obtain Majorana masses from one-loop diagrams with messengers in the loop.  Up to an overall factor, the magnitude of gaugino masses is
\begin{equation} \label{eq:1-02}
M_{\tilde{g}} \sim \frac{\alpha_s}{4 \pi}
                   \frac{\langle F_X \rangle}{\langle \phi_X \rangle} \ ,
\end{equation}
where $\alpha_s$ is the fine structure constant of the gauge group labeled by the index $s$.

It is also possible to build models with R-symmetry breaking before considering quantum corrections, or at tree level~\cite{Carpenter:2008wi, Sun:2008va}.  Such a model can be effectively described as the misalignment between $X$ and $Y$ fields.  A naive guess for gaugino masses would be
\begin{equation} \label{eq:1-03}
M_{\tilde{g}} \sim \frac{\alpha_s}{4 \pi}
                   \frac{\langle F_X \rangle}{\langle \phi_Y \rangle} \ .
\end{equation}
However, as shown in previous literature~\cite{Liu:2014ida}, the simplest type of tree-level R-symmetry breaking models has no one-loop diagram for gaugino masses with R-charge conservation at all vertexes, unless the R-charges of $X$ and $Y$ are identical and the assumption of misalignment becomes invalid.  Thus generic tree-level R-symmetry breaking models in the simplest case fail to generate gaugino masses, and are phenomenologically not favored.

In this work, we investigate the possibility of bypassing the previous no-go statement~\cite{Liu:2014ida} in more general models.  We show that with multiple $X$'s, $Y$'s and messenger fields, gaugino masses can be generated at one-loop level if the R-charge assignment of fields satisfies a certain condition.  The resulting gaugino masses from one-loop Feynman diagram calculations are always suppressed by a power of the messenger mass scale, whose exponent depends on the number of $X$ fields, although the detailed gaugino mass spectrum requires more careful calculation.  We present two simple examples to demonstrate this possibility.  Finally we discuss possible implications to model building and phenomenology.

\section{Gaugino masses from one spurion}

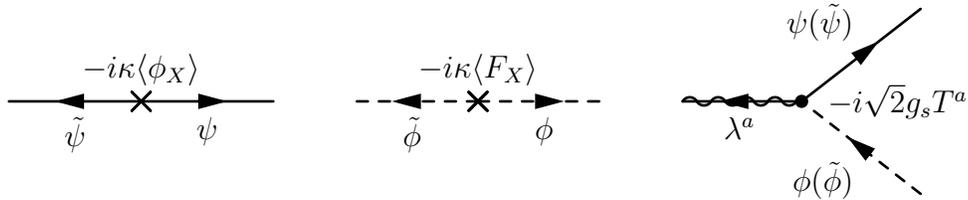
\begin{figure}
\centering 
    \begin{fmfgraph*}(100,70)
    \fmfleft{i1}
    \fmfright{o1}
    \fmf{fermion,label=$\tilde{\psi}$,label.side=left}{v1,i1}
    \fmf{fermion,label=$\psi$,label.side=right}{v1,o1}
    \fmfv{decoration.shape=cross,decoration.size=5thick,label=$- i \kappa \langle \phi_X \rangle$,label.angle=90}{v1}
    \end{fmfgraph*}
\qquad
    \begin{fmfgraph*}(100,70)
    \fmfleft{i1}
    \fmfright{o1}
    \fmf{scalar,label=$\tilde{\phi}$,label.side=left}{v1,i1}
    \fmf{scalar,label=$\phi$,label.side=right}{v1,o1}
    \fmfv{decoration.shape=cross,decoration.size=5thick,label=$- i \kappa \langle F_X \rangle$,label.angle=90}{v1}
    \end{fmfgraph*}
\qquad
    \begin{fmfgraph*}(100,70)
    \fmfleft{i1}
    \fmfright{o1,o2}
    \fmf{boson,label=$\lambda^{a}$,label.side=left}{v1,i1}
    \fmf{fermion}{v1,i1}
    \fmf{scalar,label=$\phi (\tilde{\phi})$,label.side=left}{o1,v1}
    \fmf{fermion,label=$\psi (\tilde{\psi})$,label.side=left}{v1,o2}
    \fmfv{decoration.shape=circle,decoration.size=2thick,label=$- i \sqrt{2} g_s T^a$,label.angle=0,label.dist=10thin}{v1}
    \end{fmfgraph*}
\caption{Messenger coupling vertices related to gaugino masses.} \label{fg:2-01}
\end{figure}

\begin{figure}
\centering
    \begin{fmfgraph*}(160,90)
    \fmfleft{i1}
    \fmfright{o1}
    \fmf{fermion,label=$1$,label.side=left}{v1,i1}
    \fmf{photon}{v1,i1}
    \fmf{fermion,label=$1$,label.side=right}{v2,o1}
    \fmf{photon}{v2,o1}
    \fmf{phantom,left,tension=0.5,tag=1}{v1,v2}
    \fmf{phantom,right,tension=0.5,tag=2}{v1,v2}
    \fmfdot{v1,v2}
    \fmffreeze
    \fmfipath{p[]}
    \fmfiset{p1}{vpath1(__v1,__v2)}
    \fmfiset{p2}{vpath2(__v1,__v2)}
    \fmfi{fermion,label=$r - 1$}{subpath (0,length(p1)/2) of p1}
    \fmfi{fermion,label=$- r - 1$}{subpath (length(p1),length(p1)/2) of p1}
    \fmfi{scalar,label=$r$}{subpath (length(p2)/2,0) of p2}
    \fmfi{scalar,label=$- r$}{subpath (length(p2)/2,length(p2)) of p2}
    \fmfiv{decoration.shape=cross,decoration.size=5thick,label=$\langle \phi_X^* \rangle$,label.angle=-90}{point length(p1)/2 of p1}
    \fmfiv{decoration.shape=cross,decoration.size=5thick,label=$\langle F_X \rangle$,label.angle=90}{point length(p2)/2 of p2}
    \end{fmfgraph*}
\caption{A one-loop diagram for gaugino masses, with R-charges labeled.}
\label{fg:2-02}
\end{figure}
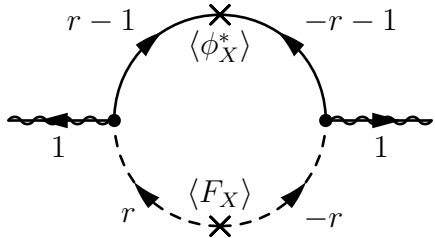

This section reviews previous results of ordinary gauge mediation, which will later be generalized to models with multiple misaligned spurions.  We start from the superfield formulation of the SUSY Lagrangian
\begin{equation}
L = L_\text{Kinetic} + [W]_{\theta \theta} + \text{h.c.}
\end{equation}
and expand it in component fields.  Ignoring the details of SUSY breaking and R-symmetry breaking mechanisms, the SUSY breaking sector is described by a spurion field $X$ with the VEV \eqref{eq:1-01} breaking both SUSY and the R-symmetry.  The R-charge of Grassmann numbers $\theta^\alpha$ is set to $1$ in the superfield formulation, so the superpotential $W$ has R-charge $2$ to make the Lagrangian R-invariant.  In simple models, $X$ usually has R-charge $2$, and couples to messenger fields through the cubic interaction in the superpotential
\begin{equation}
W = \kappa X \tilde{\Phi} \Phi \ ,
\end{equation}
where the messenger fields $\tilde{\Phi}$ and $\Phi$ have opposite R-charges $\pm r$ and are conjugate to each other as SM gauge symmetry representations.  Expanding the superpotential part of the SUSY Lagrangian in component fields, the nonzero VEV of $X$ gives quadratic vertices to messenger fields:
\begin{equation}
[\kappa X \tilde{\Phi} \Phi]_{\theta \theta} + \text{h.c.}
  = \kappa \langle \phi_X \rangle \tilde{\psi} \psi
    + \kappa \langle F_X \rangle \tilde{\phi} \phi
    + \text{h.c.} + \dotsb \ .
\end{equation}
Messengers are nontrivial representations of SM gauge symmetries, thus coupled to vector superfields $V = V^a T^a$ in the kinetic part of the Lagrangian, where $T^a$ are gauge symmetry group generators, and we use the Einstein summation convention for repeating indices throughout this work.  Expanding the kinetic part of the SUSY Lagrangian in component fields, the gauge coupling gives the cubic vertices between messenger fields and gauginos:
\begin{equation}
[\Phi^\dagger e^{2 g_s V^a T^a} \Phi]_{\theta \theta \bar{\theta} \bar{\theta}}
  = - \sqrt{2} g_s (\phi^* T^a \psi) \lambda^a
    + \text{h.c.} + \dotsb \ ,
\end{equation}
where $\lambda^a$ are the SSM gauginos $\tilde{g}$, and the gauge coupling strength $g_s$ is related to the fine structure constant $\alpha_s$ of the gauge group labeled by the index $s$:
\begin{equation}
\alpha_s = \frac{g_s^2}{4 \pi} \ .
\end{equation}
Similar vertices from gauge coupling between $\tilde{\Phi}$ and $V$ also exist.  The Feynman rules of these vertexes are shown in Figure~\ref{fg:2-01}.  Gauginos obtain Majorana masses of the magnitude \eqref{eq:1-02} from the one-loop Feynman diagram shown in Figure~\ref{fg:2-02}, where the R-charges of messenger component fields are labeled.

\begin{fmffile}{fg1}
\end{fmffile}

\begin{fmffile}{fg2}
\end{fmffile}

The messenger sector can be generalized to have more than two messenger fields, and include explicit mass terms~\cite{Cheung:2007es, Marques:2009yu}. The corresponding superpotential is
\begin{equation} \label{eq:2-01}
W = \kappa_{i j} X \tilde{\Phi}_i \Phi_j
    + m_{i j} \tilde{\Phi}_i \Phi_j \ .
\end{equation}
where each term with nonzero $\kappa_{i j}$ or $m_{i j}$ must have R-charges of fields added up to $2$.  Messenger fields with and without tilde on their symbols are conjugate to each other as SM gauge symmetry representations.  The messenger mass matrix
\begin{equation}
\mathcal{M}_{i j} = \kappa_{i j} X + m_{i j}
\end{equation}
is not diagonal and depends on the VEV of $X$.  It alters the quadratic vertices and the propagators of messengers in the loop, making the one-loop calculation more complicated.  One can instead do the calculation using the wave-function renormalization method~\cite{Giudice:1997ni}, and the resulting gaugino masses are given as in~\cite{Cheung:2007es}:
\begin{equation} \label{eq:2-02}
M_{\tilde{g}} = \frac{\alpha_s}{4 \pi}
                \langle F_X \rangle
                \left. \frac{\partial}{\partial_{X}} \right \rvert_{\langle \phi_X \rangle} \log \det \mathcal{M} \ .
\end{equation}
As a consequence of the R-symmetric superpotential \eqref{eq:2-01}, the determinant of the messenger mass matrix has the identity
\begin{equation} \label{eq:2-03}
\det \mathcal{M} = X^n G(m, \kappa) \ , \quad
n = r_X^{-1} \sum_i (2 - R(\tilde{\Phi}_i) - R(\Phi_i)) \ .
\end{equation}
Then \eqref{eq:2-02} becomes
\begin{equation} \label{eq:2-04}
M_{\tilde{g}} = \frac{\alpha}{4 \pi}
                \frac{n \langle F_X \rangle}{\langle \phi_X \rangle} \ ,
\end{equation}
which agrees with the magnitude \eqref{eq:1-02}.  In particular, a model with a classically stable pseudomodulus space always gives $n = 0$ in the identity \eqref{eq:2-03}, and the gaugino masses \eqref{eq:2-04} become zero.  To obtain a nonzero gaugino mass at one loop, the classically stable vacuum condition must be relaxed.  So the SUSY breaking pseudomodulus space must have some classically unstable region where the stabilized vacuum after including quantum corrections should avoid.  This type of classical instability also means that there must be some vacuum or runaway direction with lower potential energy than the SUSY breaking vacuum.  Thus we reached an additional argument for the necessity of metastable SUSY breaking~\cite{Komargodski:2009jf, Abel:2009ze}, which complements the argument of metastability from non-perturbative effects explicitly breaking the R-symmetry~\cite{Intriligator:2006dd, Intriligator:2007py}.  Note that both the wave-function renormalization result \eqref{eq:2-02} and the messenger mass matrix identity \eqref{eq:2-02} are based on the assumption of a single spurion $X$ breaking both SUSY and the R-symmetry.  Thus this argument of metastability from classical instability of the pseudomodulus space does not work for the following cases with multiple misaligned spurions.

\section{Gaugino masses from misaligned spurions}

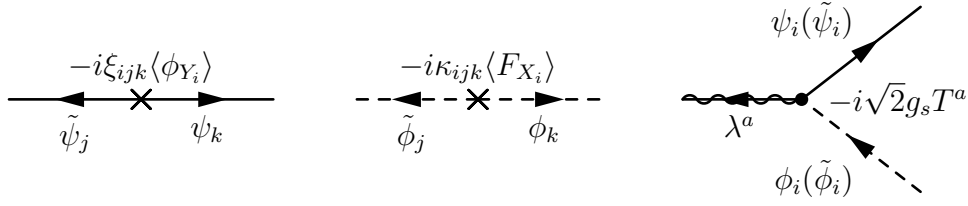
\begin{figure}
\centering 
    \begin{fmfgraph*}(100,70)
    \fmfleft{i1}
    \fmfright{o1}
    \fmf{fermion,label=$\tilde{\psi}_j$,label.side=left}{v1,i1}
    \fmf{fermion,label=$\psi_k$,label.side=right}{v1,o1}
    \fmfv{decoration.shape=cross,decoration.size=5thick,label=$- i \xi_{i j k} \langle \phi_{Y_i} \rangle$,label.angle=90}{v1}
    \end{fmfgraph*}
\qquad
    \begin{fmfgraph*}(100,70)
    \fmfleft{i1}
    \fmfright{o1}
    \fmf{scalar,label=$\tilde{\phi}_j$,label.side=left}{v1,i1}
    \fmf{scalar,label=$\phi_k$,label.side=right}{v1,o1}
    \fmfv{decoration.shape=cross,decoration.size=5thick,label=$- i \kappa_{i j k} \langle F_{X_i} \rangle$,label.angle=90}{v1}
    \end{fmfgraph*}
\qquad
    \begin{fmfgraph*}(100,70)
    \fmfleft{i1}
    \fmfright{o1,o2}
    \fmf{boson,label=$\lambda^{a}$,label.side=left}{v1,i1}
    \fmf{fermion}{v1,i1}
    \fmf{scalar,label=$\phi_i (\tilde{\phi}_i)$,label.side=left}{o1,v1}
    \fmf{fermion,label=$\psi_i (\tilde{\psi}_i)$,label.side=left}{v1,o2}
    \fmfv{decoration.shape=circle,decoration.size=2thick,label=$- i \sqrt{2} g_s T^a$,label.angle=0,label.dist=10thin}{v1}
    \end{fmfgraph*}
\caption{Messenger coupling vertices related to gaugino masses from misaligned spurions.} \label{fg:3-01}
\end{figure}

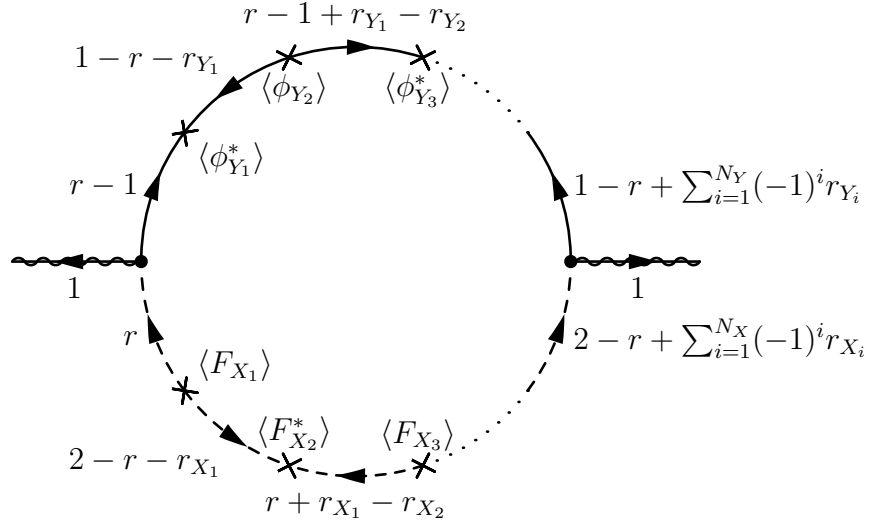
\begin{figure}
\centering
    \begin{fmfgraph*}(260,200)
    \fmfleft{i1}
    \fmfright{o1}
    \fmf{fermion,label=$1$,label.side=left}{v1,i1}
    \fmf{photon}{v1,i1}
    \fmf{fermion,label=$1$,label.side=right}{v2,o1}
    \fmf{photon}{v2,o1}
    \fmf{phantom,left,tension=0.3,tag=1}{v1,v2}
    \fmf{phantom,right,tension=0.3,tag=2}{v1,v2}
    \fmfdot{v1,v2}
    \fmffreeze
    \fmfipath{p[]}
    \fmfiset{p1}{vpath1(__v1,__v2)}
    \fmfiset{p2}{vpath2(__v1,__v2)}
    \fmfi{fermion,label=$r - 1$}{subpath (0,length(p1)/5) of p1}
    \fmfi{fermion,label=$1 - r - r_{Y_1}$}{subpath (2*length(p1)/5,length(p1)/5) of p1}
    \fmfi{fermion,label=$r - 1 + r_{Y_1} - r_{Y_2}$}{subpath (2*length(p1)/5,3*length(p1)/5) of p1}
    \fmfi{dots}{subpath (3*length(p1)/5,4*length(p1)/5) of p1}
    \fmfi{fermion,label=$1 - r + \sum_{i = 1}^{N_Y} (- 1)^i r_{Y_i}$}{subpath (length(p1),4*length(p1)/5) of p1}
    \fmfi{scalar,label=$r$}{subpath (length(p2)/5,0) of p2}
    \fmfi{scalar,label=$2 - r - r_{X_1}$}{subpath (length(p2)/5,2*length(p2)/5) of p2}
    \fmfi{scalar,label=$r + r_{X_1} - r_{X_2}$}{subpath (3*length(p2)/5,2*length(p2)/5) of p2}
    \fmfi{dots}{subpath (3*length(p2)/5,4*length(p2)/5) of p2}
    \fmfi{scalar,label=$2 - r + \sum_{i = 1}^{N_X} (- 1)^i r_{X_i}$}{subpath (4*length(p2)/5,length(p2)) of p2}
    \fmfiv{decoration.shape=cross,decoration.size=5thick,decoration.angle=54,label=$\langle \phi_{Y_1}^* \rangle$,label.angle=-36}{point length(p1)/5 of p1}
    \fmfiv{decoration.shape=cross,decoration.size=5thick,decoration.angle=18,label=$\langle \phi_{Y_2} \rangle$,label.angle=-72}{point 2*length(p1)/5 of p1}
    \fmfiv{decoration.shape=cross,decoration.size=5thick,decoration.angle=-18,label=$\langle \phi_{Y_3}^* \rangle$,label.angle=-108}{point 3*length(p1)/5 of p1}
    \fmfiv{decoration.shape=cross,decoration.size=5thick,decoration.angle=-54,label=$\langle F_{X_1} \rangle$,label.angle=36}{point length(p2)/5 of p2}
    \fmfiv{decoration.shape=cross,decoration.size=5thick,decoration.angle=-18,label=$\langle F_{X_2}^* \rangle$,label.angle=72}{point 2*length(p2)/5 of p2}
    \fmfiv{decoration.shape=cross,decoration.size=5thick,decoration.angle=18,label=$\langle F_{X_3} \rangle$,label.angle=108}{point 3*length(p2)/5 of p2}
    \end{fmfgraph*}
\caption{A possible one-loop diagram for gaugino masses from misaligned spurions, with R-charges labeled.}
\label{fg:3-02}
\end{figure}

This section generalizes the gauge mediation formulation in the previous section to tree-level R-symmetry breaking models, whose essential concept is the misalignment between the SUSY breaking spurion $X$ and the R-symmetry breaking spurion $Y$.  Instead of \eqref{eq:1-01}, we have the VEV's
\begin{equation} \label{eq:3-01}
\langle X_i \rangle = \theta^2 \langle F_{X_i} \rangle \ , \quad
\langle Y_i \rangle = \langle \phi_{Y_i} \rangle \ .
\end{equation}
Here we suppose that in a generalized model, there may be multiple $X$ and $Y$ fields with different R-charges $r(X_i)$ and $r(Y_i)$, which is a common feature of previously known examples of SUSY breaking models with tree-level R-symmetry breaking~\cite{Carpenter:2008wi, Sun:2008va, Komargodski:2009jf}.  Both $X$ and $Y$ fields couple to messenger fields through cubic interactions in the superpotential
\begin{equation} \label{eq:3-02}
W = \kappa_{i j k} X_i \tilde{\Phi}_j \Phi_k
    + \xi_{i j k} Y_i \tilde{\Phi}_j \Phi_k \ ,
\end{equation}
where each term with a nonzero coefficient $\kappa_{i j k}$ or $\xi_{i j k}$ must have R-charges of fields added up to $2$.  Messenger fields with and without tilde on their symbols are conjugate to each other as SM gauge symmetry representations.  Expanding the superpotential part of the SUSY Lagrangian in component fields, the nonzero VEV's of $X$'s and $Y$'s gives quadratic vertices to messenger fields:
\begin{align}
[\kappa_{i j k} X_i \tilde{\Phi}_j \Phi_k]_{\theta \theta} + \text{h.c.}
  &= \kappa_{i j k} \langle F_{X_i} \rangle \tilde{\phi}_j \phi_k
     + \text{h.c.} + \dotsb \ , \\
[\xi_{i j k} Y_i \tilde{\Phi}_j \Phi_k]_{\theta \theta} + \text{h.c.}
  &= \xi_{i j k} \langle \phi_{Y_i} \rangle \tilde{\psi}_j \psi_k
     + \text{h.c.} + \dotsb \ ,
\end{align}
and the gauge coupling in the kinetic part of the SUSY Lagrangian gives the cubic vertices between messenger fields and gauginos:
\begin{equation}
[\Phi_i^\dagger e^{2 g V^a T^a} \Phi_i]_{\theta \theta \bar{\theta} \bar{\theta}}
  = - \sqrt{2} g (\phi_i^* T^a \psi_i) \lambda^a
    + \text{h.c.} + \dotsb \ .
\end{equation}
Similar vertices from gauge coupling between $\tilde{\Phi}$ and $V$ also exist.  The Feynman rules of these vertexes are shown in Figure~\ref{fg:3-01}.  Note that a quadratic vertex may mix messenger fields with different indices, but a cubic vertex always connects a gaugino with fermion and scalar components of the same messenger superfield.  One can insert multiple quadratic vertices into both the fermion part and the scalar part of the loop, and gauginos may obtain Majorana masses from the one-loop Feynman diagram shown in Figure~\ref{fg:3-02}.  If the diagram does exist, R-charge conservation should be satisfied at each vertex.  Thus R-charges of messengers can be determined by $r(X_i)$, $r(Y_i)$ and $r$, the R-charge of the first messenger connecting to the gaugino on the left of the loop diagram.  The resulting R-charges of messenger component fields are labeled in Figure~\ref{fg:3-02}.  Then R-charge conservation at the vertex connecting messengers to the gaugino on the right of the loop diagram leads to the condition
\begin{equation} \label{eq:3-03}
\sum_{i = 1}^{N_X} (- 1)^i r(X_i) = \sum_{i = 1}^{N_Y} (- 1)^i r(Y_i) \ ,
\end{equation}
where the sum is over the $\langle F_X \rangle$ vertices or the $\langle \phi_Y \rangle$ vertices in the loop diagram.  Noticing the arrow directions of messenger field propagators and pairs of messenger with and without tilde connected to quadratic vertices, both the fermion part and the scalar part of the loop diagram in Figure~\ref{fg:3-02} must have odd numbers of vertices inserted.  Thus both $N_X$ and $N_Y$ must be odd numbers.

\begin{fmffile}{fg3}
\end{fmffile}

\begin{fmffile}{fg4}
\end{fmffile}

To obtain the magnitude of gaugino masses from the one-loop Feynman diagram shown in Figure~\ref{fg:3-02},  we simplify the coefficients of cubic interactions in the superpotential \eqref{eq:3-02} to just two universal parameters $\kappa = \kappa_{i j k}$ and $\xi = \xi_{i j k}$.  We assume for simplicity that all VEV's of $X$'s and $Y$'s are at the universal scales $\langle F_X \rangle$ and $\langle \phi_Y \rangle$, and all messengers obtain masses at the universal scale $M \sim \langle \phi_Y \rangle$.  Notice also that all Feynman rules of quadratic vertices can be set to real by proper complex phase rotations of messenger fields.  Now we are interested in the regime $\langle F_X \rangle \ll M$ so that an effective field theory calculation can be done to give soft masses from the SUSY breaking effect.  Using the two-component spinor techniques for quantum field theory developed in~\cite{Dreiner:2008tw}, a straightforward calculation of the one-loop diagram in Figure~\ref{fg:3-02} gives
\begin{equation} \label{eq:3-04}
\begin{split}
- i M_{\tilde g} (p)
  &= - 2 g_s^2
     (\kappa \langle F_X \rangle)^{N_X}
     (\xi \langle \phi_Y \rangle)^{N_Y}
     \int \frac{\ud^4 k}{(2 \pi)^4}
          \frac{(k_\mu \sigma^\mu k_\nu \bar{\sigma}^\nu)^{(N_Y + 1)/2}}
               {((k + p)^2 - M^2)^{N_X + 1} (k^2 - M^2)^{N_Y + 1}}\\
  &= - 2 g_s^2
     (\kappa \langle F_X \rangle)^{N_X}
     (\xi \langle \phi_Y \rangle)^{N_Y}
     \int \frac{\ud^4 k}{(2 \pi)^4}
          \frac{k^{N_Y + 1}}
               {((k + p)^2 - M^2)^{N_X + 1} (k^2 - M^2)^{N_Y + 1}} \ .
\end{split}
\end{equation}
Following the standard procedure in quantum field theory textbooks~\cite{Peskin:1995ev}, the integration can be done by introducing a Feynman parameter $x$ and performing a Wick rotation:
\begin{equation}
\begin{split}
&\int \frac{\ud^4 k}{(2 \pi)^4}
      \frac{k^{N_Y + 1}}
           {((k + p)^2 - M^2)^{N_X + 1} (k^2 - M^2)^{N_Y + 1}}\\
  &\qquad = \frac{(N_X + N_Y + 1)!}{N_X! \, N_Y!}
            \int_0^1 \ud x
            \int \frac{\ud^4 k}{(2 \pi)^4}
                 \frac{x^{N_X} (1 - x)^{N_Y} k^{N_Y + 1}}
                      {(k^2 + 2 x k \cdot p + x p^2 - M^2)^{N_X + N_Y + 2}}\\
  &(l = k + x p, \ \Delta = M^2 - x (1 - x) p^2)\\
  &\qquad = \frac{(N_X + N_Y + 1)!}{N_X! \, N_Y!}
            \int_0^1 \ud x
            \int \frac{\ud^4 l}{(2 \pi)^4}
                 \frac{x^{N_X} (1 - x)^{N_Y} (l - x p)^{N_Y + 1}}
                      {(l^2 - \Delta)^{N_X + N_Y + 2}}\\
  &\qquad = \sum_{a = 0}^{(N_Y + 1) / 2}
            \frac{(N_X + N_Y + 1)! \, (N_Y + 1)}
                 {N_X! \, (N_Y + 1 - 2 a)! \, (2 a)!} \times\\
  &\hspace{7em} \times
            \int_0^1 \ud x \,
                     x^{N_X} (1 - x)^{N_Y} (x p)^{N_Y + 1 - 2 a}
            \int \frac{\ud^4 l}{(2 \pi)^4}
                 \frac{l^{2 a}}
                      {(l^2 - \Delta)^{N_X + N_Y + 2}}\\
  &(l_E^0 = - i l^0)\\
  &\qquad = \sum_{a = 0}^{(N_Y + 1) / 2}
            i (- 1)^a
            \frac{(N_X + N_Y + 1)! \, (N_Y + 1)}
                 {N_X! \, (N_Y + 1 - 2 a)! \, (2 a)!} \times\\
  &\hspace{7em} \times
            \int_0^1 \ud x \,
                     x^{N_X} (1 - x)^{N_Y} (x p)^{N_Y + 1 - 2 a}
            \int \frac{\ud^4 l_E}{(2 \pi)^4}
                 \frac{l_E^{2 a}}
                      {(l_E^2 + \Delta)^{N_X + N_Y + 2}}\\
  &\qquad = \sum_{a = 0}^{(N_Y + 1) / 2}
            \frac{i (- 1)^a}{(4 \pi)^2}
            \frac{(N_X + N_Y - a - 1)! \, (N_Y + 1) (a + 1)!}
                 {N_X! \, (N_Y + 1 - 2 a)! \, (2 a)!} \times\\
  &\hspace{7em} \times
            \int_0^1 \ud x \,
                     \frac{x^{N_X} (1 - x)^{N_Y} (x p)^{N_Y + 1 - 2 a}}
                          {\Delta^{N_X + N_Y - a}} \ .
\end{split}
\end{equation}
Plugging this result into \eqref{eq:3-04} and setting $p^2 = m^2$, we obtain the one-loop contribution to gaugino masses with the awkward integration and sum expressions.  Note that all integrations over the Feynman parameter $x$ in the sum are finite.  Neglecting an overall factor, it is easy to get the magnitude of gaugino masses:
\begin{equation} \label{eq:3-05}
M_{\tilde{g}} \sim \frac{\alpha_s}{4 \pi}
                   \frac{\langle F_X \rangle^{N_X} \langle \phi_Y \rangle)^{N_Y}}
                        {M^{2 N_X + N_Y - 1}}
              \sim \frac{\alpha_s}{4 \pi}
                   \frac{\langle F_X \rangle^{N_X}}{M^{2 N_X - 1}} \ ,
\end{equation}
which is always suppressed by a power of the messenger mass scale.

For models with only one $\langle F_X \rangle$ vertex and one $\langle \phi_Y \rangle$ vertex inserted in the loop of Figure~\ref{fg:3-02}, it seems that \eqref{eq:3-05} gives the magnitude of gaugino masses
\begin{equation}
M_{\tilde{g}} \sim \frac{\alpha_s}{4 \pi}
                   \frac{\langle F_X \rangle \langle \phi_Y \rangle}{M^2}
              \sim \frac{\alpha_s}{4 \pi}
                   \frac{\langle F_X \rangle}{M} \ .
\end{equation}
However, the condition \eqref{eq:3-03} in this case means $r(X) = r(Y)$.  The R-symmetry requires $X$ and $Y$ to share the same set of coupling terms with messengers.  There is no distinction between $X$ and $Y$ in terms of R-charges.  Therefore the misaligned VEV's in \eqref{eq:3-01} become not generic, and the scalar component of $X$ may also obtains a VEV to break the R-symmetry in a generic model.  This is just the no-go statement obtained in~\cite{Liu:2014ida}:  The simplest tree-level R-symmetry breaking model, if successfully generates gaugino masses in gauge mediation, is generically accompanied by same magnitude of loop-level R-symmetry breaking which also contributes to gaugino masses.  To generate gaugino masses from misaligned SUSY breaking and R-symmetry breaking spurions in a generic model, it is necessary to have multiple $\langle F_X \rangle$'s or multiple $\langle \phi_Y \rangle$'s inserted in the loop of Figure~\ref{fg:3-02}, satisfying the condition \eqref{eq:3-03}.

\section{Examples}

This section presents simple examples to demonstrate the possibility of generating gaugino masses with multiple spurions satisfying the condition \eqref{eq:3-03}.  Noticing that both $N_X$ and $N_Y$ must be odd numbers, the simplest model has either three $\langle F_X \rangle$ vertices or three $\langle \phi_Y \rangle$ vertices inserted in the loop.  One can further simplify the model by allowing different vertices to come from the same $X$ or $Y$ as long as they are not next to each other.  Therefore the simplest model may be constructed with one $X$ field and two $Y$ fields, or two $X$ fields and one $Y$ field.

\subsection{The $1 X 2 Y$ model}

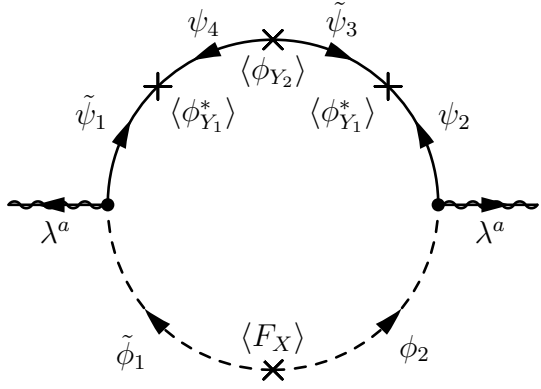
\begin{figure}
\centering
    \begin{fmfgraph*}(200,140)
    \fmfleft{i1}
    \fmfright{o1}
    \fmf{fermion,label=$\lambda^a$,label.side=left}{v1,i1}
    \fmf{photon}{v1,i1}
    \fmf{fermion,label=$\lambda^a$,label.side=right}{v2,o1}
    \fmf{photon}{v2,o1}
    \fmf{phantom,left,tension=0.3,tag=1}{v1,v2}
    \fmf{phantom,right,tension=0.3,tag=2}{v1,v2}
    \fmfdot{v1,v2}
    \fmffreeze
    \fmfipath{p[]}
    \fmfiset{p1}{vpath1(__v1,__v2)}
    \fmfiset{p2}{vpath2(__v1,__v2)}
    \fmfi{fermion,label=$\tilde{\psi}_1$}{subpath (0,length(p1)/4) of p1}
    \fmfi{fermion,label=$\psi_4$}{subpath (length(p1)/2,length(p1)/4) of p1}
    \fmfi{fermion,label=$\tilde{\psi}_3$}{subpath (length(p1)/2,3*length(p1)/4) of p1}
    \fmfi{fermion,label=$\psi_2$}{subpath (length(p1),3*length(p1)/4) of p1}
    \fmfi{scalar,label=$\tilde{\phi}_1$}{subpath (length(p2)/2,0) of p2}
    \fmfi{scalar,label=$\phi_2$}{subpath (length(p2)/2,length(p2)) of p2}
    \fmfiv{decoration.shape=cross,decoration.size=5thick,decoration.angle=45,label=$\langle \phi_{Y_1}^* \rangle$,label.angle=-45}{point length(p1)/4 of p1}
    \fmfiv{decoration.shape=cross,decoration.size=5thick,label=$\langle \phi_{Y_2} \rangle$,label.angle=-90}{point length(p1)/2 of p1}
    \fmfiv{decoration.shape=cross,decoration.size=5thick,decoration.angle=-45,label=$\langle \phi_{Y_1}^* \rangle$,label.angle=-135}{point 3*length(p1)/4 of p1}
    \fmfiv{decoration.shape=cross,decoration.size=5thick,label=$\langle F_X \rangle$,label.angle=90}{point length(p2)/2 of p2}
    \end{fmfgraph*}
\caption{A one-loop diagram for gaugino masses in the $1 X 2 Y$ model.}
\label{fg:4-01}
\end{figure}

In the first example, we have one SUSY breaking spurion $X$, and two R-symmetry breaking spurions $Y_1$ and $Y_2$.  The couplings between $X$, $Y$'s and messenger fields are described by the superpotential
\begin{equation}
W = \kappa X \tilde{\Phi}_1 \Phi_2
    + \xi_1 Y_1 \tilde{\Phi}_1 \Phi_4
    + \xi'_1 Y_1 \tilde{\Phi}_3 \Phi_2
    + \xi_2 Y_2 \tilde{\Phi}_3 \Phi_4 \ ,
\end{equation}
which is compatible with the R-charge assignment:
\begin{equation}
\begin{gathered}
r(Y_1) = r_1 \ , \quad
r(Y_2) = r_2 \ , \quad
r(X) = 2 r_1 - r_2 \ , \quad
r(\tilde{\Phi}_1) = r \ ,\\
r(\Phi_2) = 2 - r - 2 r_1 + r_2 \ , \quad
r(\tilde{\Phi}_3) = r + r_1 - r_2 \ , \quad
r(\Phi_4) = 2 - r - r_1 \ ,
\end{gathered}
\end{equation}
where the free parameters $r$, $r_1$ and $r_2$ may be fixed by some mechanism in the ultraviolet theory, such as anomaly cancellation when the R-symmetry is gauged in supergravity.  According to \eqref{eq:3-05}, the one-loop Feynmann diagram shown in Figure~\ref{fg:4-01} gives the magnitude of gaugino masses
\begin{equation} \label{eq:4-01}
M_{\tilde{g}} \sim \frac{\alpha_s}{4 \pi}
                   \frac{\langle F_X \rangle \langle \phi_Y \rangle^3}{M^4}
              \sim \frac{\alpha_s}{4 \pi}
                   \frac{\langle F_X \rangle}{M} \ .
\end{equation}

\begin{fmffile}{fg5}
\end{fmffile}

\begin{figure}
\centering
    \begin{fmfgraph*}(180,120)
    \fmfleft{i1}
    \fmfright{o1}
    \fmf{fermion,label=$\lambda^a$,label.side=left}{v1,i1}
    \fmf{photon}{v1,i1}
    \fmf{fermion,label=$\lambda^a$,label.side=right}{v2,o1}
    \fmf{photon}{v2,o1}
    \fmf{phantom,left,tension=0.3,tag=1}{v1,v2}
    \fmf{phantom,right,tension=0.3,tag=2}{v1,v2}
    \fmfdot{v1,v2}
    \fmffreeze
    \fmfipath{p[]}
    \fmfiset{p1}{vpath1(__v1,__v2)}
    \fmfiset{p2}{vpath2(__v1,__v2)}
    \fmfi{fermion,label=$\tilde{\psi}_1$}{subpath (0,length(p1)/4) of p1}
    \fmfi{fermion,label=$\psi_4$}{subpath (length(p1)/2,length(p1)/4) of p1}
    \fmfi{fermion,label=$\tilde{\psi}_3$}{subpath (length(p1)/2,3*length(p1)/4) of p1}
    \fmfi{fermion,label=$\psi_2$}{subpath (length(p1),3*length(p1)/4) of p1}
    \fmfi{scalar,label=$\tilde{\phi}_1$}{subpath (length(p2)/2,0) of p2}
    \fmfi{scalar,label=$\phi_2$}{subpath (length(p2)/2,length(p2)) of p2}
    \fmfiv{decoration.shape=cross,decoration.size=5thick,decoration.angle=45,label=$m^*$,label.angle=-45}{point length(p1)/4 of p1}
    \fmfiv{decoration.shape=cross,decoration.size=5thick,label=$\langle \phi_{Y} \rangle$,label.angle=-90}{point length(p1)/2 of p1}
    \fmfiv{decoration.shape=cross,decoration.size=5thick,decoration.angle=-45,label=$m'^*$,label.angle=-135}{point 3*length(p1)/4 of p1}
    \fmfiv{decoration.shape=cross,decoration.size=5thick,label=$\langle F_X \rangle$,label.angle=90}{point length(p2)/2 of p2}
    \end{fmfgraph*}
\qquad
    \begin{fmfgraph*}(180,120)
    \fmfleft{i1}
    \fmfright{o1}
    \fmf{fermion,label=$\lambda^a$,label.side=left}{v1,i1}
    \fmf{photon}{v1,i1}
    \fmf{fermion,label=$\lambda^a$,label.side=right}{v2,o1}
    \fmf{photon}{v2,o1}
    \fmf{phantom,left,tension=0.3,tag=1}{v1,v2}
    \fmf{phantom,right,tension=0.3,tag=2}{v1,v2}
    \fmfdot{v1,v2}
    \fmffreeze
    \fmfipath{p[]}
    \fmfiset{p1}{vpath1(__v1,__v2)}
    \fmfiset{p2}{vpath2(__v1,__v2)}
    \fmfi{fermion,label=$\tilde{\psi}_1$}{subpath (0,length(p1)/4) of p1}
    \fmfi{fermion,label=$\psi_4$}{subpath (length(p1)/2,length(p1)/4) of p1}
    \fmfi{fermion,label=$\tilde{\psi}_3$}{subpath (length(p1)/2,3*length(p1)/4) of p1}
    \fmfi{fermion,label=$\psi_2$}{subpath (length(p1),3*length(p1)/4) of p1}
    \fmfi{scalar,label=$\tilde{\phi}_1$}{subpath (length(p2)/2,0) of p2}
    \fmfi{scalar,label=$\phi_2$}{subpath (length(p2)/2,length(p2)) of p2}
    \fmfiv{decoration.shape=cross,decoration.size=5thick,decoration.angle=45,label=$\langle \phi_{Y}^* \rangle$,label.angle=-45}{point length(p1)/4 of p1}
    \fmfiv{decoration.shape=cross,decoration.size=5thick,label=$m$,label.angle=-90}{point length(p1)/2 of p1}
    \fmfiv{decoration.shape=cross,decoration.size=5thick,decoration.angle=-45,label=$\langle \phi_{Y}^* \rangle$,label.angle=-135}{point 3*length(p1)/4 of p1}
    \fmfiv{decoration.shape=cross,decoration.size=5thick,label=$\langle F_X \rangle$,label.angle=90}{point length(p2)/2 of p2}
    \end{fmfgraph*}
\caption{One-loop diagrams for gaugino masses in two variations of the $1 X 2 Y$ model.}
\label{fg:4-02}
\end{figure}
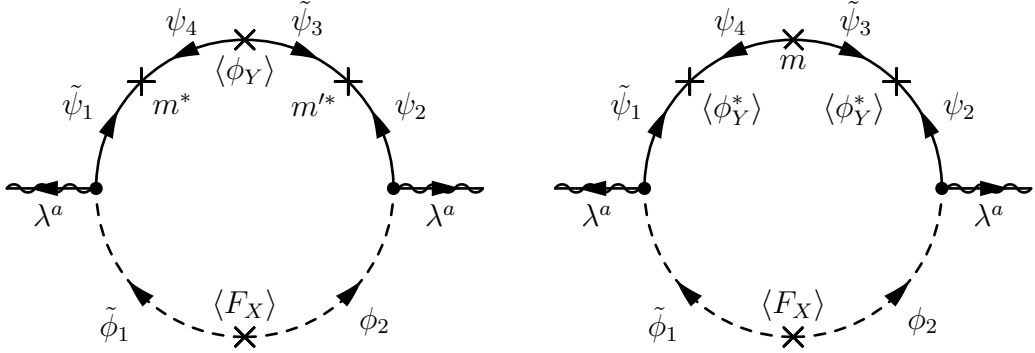

If there are explicit messenger mass terms in a model, the mass coefficients can be viewed as VEV's of some $Y$ fields with R-charge $0$ which do not violate the R-symmetry of the model.  So some $\langle \phi_{Y_i} \rangle$ vertices in the loop of Figure~\ref{fg:4-01} can be replaced by explicit mass insertions, and the model may be built with only one $X$ field and one $Y$ field.  If we replace the two $\langle \phi_{Y_1} \rangle$ vertices with masses in the loop, we have the superpotential
\begin{equation}
W = \kappa X \tilde{\Phi}_1 \Phi_2 + m \tilde{\Phi}_1 \Phi_4 + m' \tilde{\Phi}_3 \Phi_2 + \xi Y \tilde{\Phi}_3 \Phi_4
\end{equation}
with the R-charge assignment
\begin{equation}
\begin{gathered}
r(Y) = r_1 \ , \quad
r(X) = - r_1 \ , \quad
r(\tilde{\Phi}_1) = r \ ,\\
r(\Phi_2) = 2 - r + r_1 \ , \quad
r(\tilde{\Phi}_3) = r - r_1 \ , \quad
r(\Phi_4) = 2 - r \ .
\end{gathered}
\end{equation}
We can also replace the one $\langle \phi_{Y_2} \rangle$ vertex with a mass in the loop, then we have the superpotential
\begin{equation}
W = \kappa X \tilde{\Phi}_1 \Phi_2 + \xi Y \tilde{\Phi}_1 \Phi_4 + \xi' Y \tilde{\Phi}_3 \Phi_2 + m \tilde{\Phi}_3 \Phi_4
\end{equation}
with the R-charge assignment
\begin{equation}
\begin{gathered}
r(Y) = r_1 \ , \quad
r(X) = 2 r_1 \ , \quad
r(\tilde{\Phi}_1) = r \ ,\\
r(\Phi_2) = 2 - r - 2 r_1 \ , \quad
r(\tilde{\Phi}_3) = r + r_1 \ , \quad
r(\Phi_4) = 2 - r - r_1 \ .
\end{gathered}
\end{equation}
The one-loop Feynman diagrams for gaugino masses in these two variations of the $1 X 2 Y$ model are shown in Figure~\ref{fg:4-02}.  Both of them give the same magnitudes of gaugino masses as \eqref{eq:4-01} if we assume $\langle \phi_Y \rangle$ and the explicit messenger masses are at the universal scale $M$.

\begin{fmffile}{fg6}
\end{fmffile}

\subsection{The $2 X 1 Y$ model}

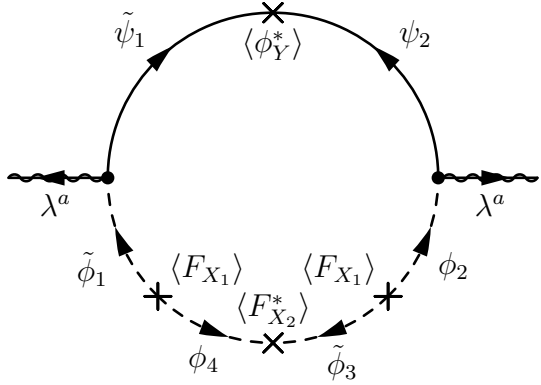
\begin{figure}
\centering
    \begin{fmfgraph*}(200,140)
    \fmfleft{i1}
    \fmfright{o1}
    \fmf{fermion,label=$\lambda^a$,label.side=left}{v1,i1}
    \fmf{photon}{v1,i1}
    \fmf{fermion,label=$\lambda^a$,label.side=right}{v2,o1}
    \fmf{photon}{v2,o1}
    \fmf{phantom,left,tension=0.3,tag=1}{v1,v2}
    \fmf{phantom,right,tension=0.3,tag=2}{v1,v2}
    \fmfdot{v1,v2}
    \fmffreeze
    \fmfipath{p[]}
    \fmfiset{p1}{vpath1(__v1,__v2)}
    \fmfiset{p2}{vpath2(__v1,__v2)}
    \fmfi{fermion,label=$\tilde{\psi}_1$}{subpath (0,length(p1)/2) of p1}
    \fmfi{fermion,label=$\psi_2$}{subpath (length(p1),length(p1)/2) of p1}
    \fmfi{scalar,label=$\tilde{\phi}_1$}{subpath (length(p2)/4,0) of p2}
    \fmfi{scalar,label=$\phi_4$}{subpath (length(p2)/4,length(p2)/2) of p2}
    \fmfi{scalar,label=$\tilde{\phi}_3$}{subpath (3*length(p2)/4,length(p2)/2) of p2}
    \fmfi{scalar,label=$\phi_2$}{subpath (3*length(p2)/4,length(p2)) of p2}
    \fmfiv{decoration.shape=cross,decoration.size=5thick,label=$\langle \phi_{Y}^* \rangle$,label.angle=-90}{point length(p1)/2 of p1}
    \fmfiv{decoration.shape=cross,decoration.size=5thick,decoration.angle=-45,label=$\langle F_{X_1} \rangle$,label.angle=45}{point length(p2)/4 of p2}
    \fmfiv{decoration.shape=cross,decoration.size=5thick,label=$\langle F_{X_2}^* \rangle$,label.angle=90}{point length(p2)/2 of p2}
    \fmfiv{decoration.shape=cross,decoration.size=5thick,decoration.angle=45,label=$\langle F_{X_1} \rangle$,label.angle=135}{point 3*length(p2)/4 of p2}
    \end{fmfgraph*}
\caption{A one-loop diagram for gaugino masses in the $2 X 1 Y$ model.}
\label{fg:4-03}
\end{figure}

In the second example, we have two SUSY breaking spurions $X_1$ and $X_2$, and one R-symmetry breaking spurion $Y$.  The couplings between $X$'s, $Y$ and messenger fields are described by the superpotential
\begin{equation}
W = \kappa_1 X_1 \tilde{\Phi}_1 \Phi_4 + \kappa'_1 X_1 \tilde{\Phi}_3 \Phi_2 + \kappa_2 X_2 \tilde{\Phi}_3 \Phi_4 + \xi Y \tilde{\Phi}_1 \Phi_2 \ ,
\end{equation}
which is compatible with the R-charge assignment:
\begin{equation}
\begin{gathered}
r(X_1) = r_1 \ , \quad
r(X_2) = r_2 \ , \quad
r(Y) = 2 r_1 - r_2 \ , \quad
r(\tilde{\Phi}_1) = r \ ,\\
r(\Phi_2) = 2 - r - 2 r_1 + r_2 \ , \quad
r(\tilde{\Phi}_3) = r + r_1 - r_2 \ , \quad
r(\Phi_4) = 2 - r - r_1 \ ,
\end{gathered}
\end{equation}
where the free parameters $r$, $r_1$ and $r_2$ may be fixed by some mechanism in the ultraviolet theory.  According to \eqref{eq:3-05}, the one-loop Feynmann diagram shown in Figure~\ref{fg:4-03} gives the magnitude of gaugino masses
\begin{equation}
M_{\tilde{g}} \sim \frac{\alpha_s}{4 \pi}
                   \frac{\langle F_X \rangle^3 \langle \phi_Y \rangle}{M^6}
              \sim \frac{\alpha_s}{4 \pi}
                   \frac{\langle F_X \rangle^3}{M^5} \ ,
\end{equation}
which is suppressed by a higher power of $M$ because of multiple $\langle F_X \rangle$ insertions in the loop of Figure~\ref{fg:4-03}. 

\begin{fmffile}{fg7}
\end{fmffile}

\section{Conclusion}

In this work, we have shown that it is possible to generate gaugino masses, using multiple SUSY breaking and R-symmetry breaking spurions with a proper R-charge arrangement satisfying the condition \eqref{eq:3-03}.  The magnitude of resulting gaugino masses \eqref{eq:3-05} is suppressed by a power of the messenger mass scale $M$.  The exponent of $M$ in the denominator depends on $N_X$, the number of $\langle F_X \rangle$ quadratic vertices inserted in the scalar part of the loop diagram shown in Figure~\ref{fg:3-02}.  In particular, models with one SUSY breaking spurion $X$ always result in gaugino masses of the magnitude $M_{\tilde{g}} \sim \frac{\alpha_s}{4 \pi} \frac{\langle F_X \rangle}{M}$, provided that all VEV's of $Y$'s and all messenger masses are at the universal scale $M$.  In realistic models, there may be some hierarchy between VEV's of different $X$'s and $Y$'s as well as coefficients of different cubic interactions in the superpotential.  Messenger masses must be determined from eigenvalues of the messenger mass matrix, and the mass eigenstates are generally not aligned with the R-charge eigenstates of messengers.  There may also be more than one combinations of $X$'s and $Y$'s satisfying the condition \eqref{eq:3-03}, corresponding to multiple different loop diagrams which contribute to the gaugino masses.  Therefore the actual gaugino masses require more careful calculation depending on these details of the model. 

The misalignment between SUSY breaking and R-symmetry breaking spurions may arise not only in tree-level R-symmetry breaking models~\cite{Carpenter:2008wi, Sun:2008va, Komargodski:2009jf}, but also in models with separate SUSY breaking and R-symmetry breaking sectors.  The recently found counterexamples to the Nelson-Seiberg theorem shows such possibility of R-symmetry breaking spurions which do not break SUSY~\cite{Sun:2019bnd, Amariti:2020lvx, Sun:2021svm, Brister:2022rrz}.  So there could be more freedom to manipulate the mass spectrum of gauginos as well as other SUSY breaking soft terms, by independently adjusting the SUSY breaking and R-symmetry breaking effects in different sectors.  Gaugino masses are suppressed by a higher power of the messenger mass scale $M$ in models which require multiple $\langle F_X \rangle$ insertions in the one-loop diagrams to satisfy the condition \eqref{eq:3-03}.  On the other hand, sfermion masses from two-loop diagrams do not suffer from such suppression, thus may remain heavy.  This splitting of sparticle masses gives rise to the split-SUSY scenario~\cite{Arkani-Hamed:2004ymt, Giudice:2004tc, Arkani-Hamed:2004zhs}, which has gained attention and development in the era of null discoveries from LHC and dark matter searches~\cite{Harigaya:2013asa, Wang:2013rba, Wang:2015mea, Ahmed:2019xon}.

The multiple spurions in our assumption could be just artifacts of the R-symmetry, that one single SUSY breaking or R-symmetry breaking spurion is a linear combination of spurions with different R-charges.  This is the usual situation of tree-level R-symmetry breaking models~\cite{Carpenter:2008wi, Sun:2008va, Komargodski:2009jf} with a single sector possessing both SUSY breaking and R-symmetry breaking.  An alternative possibility is to have multiple sectors, and each sector independently breaks SUSY or the R-symmetry.  The multiple decoupled SUSY breaking sectors correspondingly yield multiple massless goldstini~\cite{Cheung:2010mc}, which could lead to interesting phenomenology such as models for goldstini dark matter\cite{Cheng:2010mw, Cao:2020oxq}.

\section*{Acknowledgement}

We thank Michael Dine, Feihu Liu, Muyang Liu and Dimitri Polyakov for helpful discussions.  TL is supported in part by the National Key Research and Development Program of China, Grant No.\ 2020YFC2201504, by the National Natural Science Foundation of China, Grant No.\ 11875062, No.\ 11947302, No.\ 12047503,
and No.\ 12275333, by the Key Research Program of the Chinese Academy of Sciences, Grant No.\ XDPB15, by the Scientific Instrument Developing Project of the Chinese Academy of Sciences, Grant No.\ YJKYYQ20190049, and by the International Partnership Program of Chinese Academy of Sciences for Grand Challenges, Grant No.\ 112311KYSB20210012.  ZS is supported by the National Natural Science Foundation of China, Grant No.\ 11305110.


\begin{thebibliography}{99}

\bibitem{Martin:1997ns}
S.~P.~Martin,
``A Supersymmetry primer,''
Adv. Ser. Direct. High Energy Phys. \textbf{18} (1998), 1-98
[arXiv:hep-ph/9709356 [hep-ph]].

\bibitem{Giudice:1998bp}
G.~F.~Giudice and R.~Rattazzi,
``Theories with gauge mediated supersymmetry breaking,''
Phys. Rept. \textbf{322} (1999), 419-499
[arXiv:hep-ph/9801271 [hep-ph]].

\bibitem{Meade:2008wd}
P.~Meade, N.~Seiberg and D.~Shih,
``General Gauge Mediation,''
Prog. Theor. Phys. Suppl. \textbf{177} (2009), 143-158
[arXiv:0801.3278 [hep-ph]].

\bibitem{Kitano:2010fa}
R.~Kitano, H.~Ooguri and Y.~Ookouchi,
``Supersymmetry Breaking and Gauge Mediation,''
Ann. Rev. Nucl. Part. Sci. \textbf{60} (2010), 491-511
[arXiv:1001.4535 [hep-th]].

\bibitem{ParticleDataGroup:2022pth}
R.~L.~Workman \textit{et al.} [Particle Data Group],
``Review of Particle Physics,''
PTEP \textbf{2022} (2022), 083C01.

\bibitem{Intriligator:2007cp}
K.~A.~Intriligator and N.~Seiberg,
Class. Quant. Grav. \textbf{24} (2007), S741-S772
doi:10.1088/0264-9381/24/21/S02
[arXiv:hep-ph/0702069 [hep-ph]].

\bibitem{Wess:1973kz}
J.~Wess and B.~Zumino,
``A Lagrangian Model Invariant Under Supergauge Transformations,''
Phys. Lett. B \textbf{49} (1974), 52.

\bibitem{Wess:1974jb}
J.~Wess and B.~Zumino,
``Supergauge Invariant Extension of Quantum Electrodynamics,''
Nucl. Phys. B \textbf{78} (1974), 1.

\bibitem{Nelson:1993nf}
A.~E.~Nelson and N.~Seiberg,
``R symmetry breaking versus supersymmetry breaking,''
Nucl. Phys. B \textbf{416} (1994), 46-62
[arXiv:hep-ph/9309299 [hep-ph]].

\bibitem{Sun:2011fq}
Z.~Sun,
``Low energy supersymmetry from R-symmetries,''
Phys. Lett. B \textbf{712} (2012), 442-444
[arXiv:1109.6421 [hep-th]].

\bibitem{Kang:2012fn}
Z.~Kang, T.~Li and Z.~Sun,
``The Nelson-Seiberg theorem revised,''
JHEP \textbf{12} (2013), 093
[arXiv:1209.1059 [hep-th]].

\bibitem{Sun:2019bnd}
Z.~Sun, Z.~Tan and L.~Yang,
``A counterexample to the Nelson-Seiberg theorem,''
JHEP \textbf{10} (2020), 072
[arXiv:1904.09589 [hep-th]].

\bibitem{Amariti:2020lvx}
A.~Amariti and D.~Sauro,
``On the Nelson-Seiberg theorem: generalizations and counter-examples,''
[arXiv:2005.02076 [hep-th]].

\bibitem{Li:2020wdk}
Z.~Li and Z.~Sun,
``The Nelson-Seiberg theorem generalized with nonpolynomial superpotentials,''
Adv. High Energy Phys. \textbf{2020} (2020), 3701943
[arXiv:2006.00538 [hep-th]].

\bibitem{Sun:2021svm}
Z.~Sun, Z.~Tan and L.~Yang,
``A sufficient condition for counterexamples to the Nelson-Seiberg theorem,''
JHEP \textbf{07} (2021), 175
[arXiv:2106.08879 [hep-th]].

\bibitem{Li:2021ydn}
Z.~Li and Z.~Sun,
``Nonexistence of supersymmetry breaking counterexamples to the Nelson-Seiberg theorem,''
JHEP \textbf{10} (2021), 170
[arXiv:2107.09943 [hep-th]].

\bibitem{Brister:2021xca}
J.~Brister, Z.~Sun and G.~Yang,
``A formal notion of genericity and term-by-term vanishing superpotentials at supersymmetric vacua from R-symmetric Wess-Zumino models,''
JHEP \textbf{12} (2021), 199
[arXiv:2111.09570 [hep-th]].

\bibitem{Brister:2022rrz}
J.~Brister and Z.~Sun,
``Novel counterexample to the Nelson-Seiberg theorem,''
Phys. Rev. D \textbf{106} (2022) no.8, 8
[arXiv:2203.05464 [hep-th]].

\bibitem{Brister:2022vsz}
J.~Brister, S.~Kou, Z.~Li and Z.~Sun,
``A brute-force search for R-symmetric Wess-Zumino models,''
[arXiv:2204.05767 [hep-th]].

\bibitem{Sun:2022xdl}
Z.~Sun,
``Supersymmetry and R-symmetries in Wess-Zumino models: properties and model dataset construction,''
[arXiv:2207.13933 [hep-th]].

\bibitem{Ray:2006wk}
S.~Ray,
``Some properties of meta-stable supersymmetry-breaking vacua in Wess-Zumino models,''
Phys. Lett. B \textbf{642} (2006), 137-141
[arXiv:hep-th/0607172 [hep-th]].

\bibitem{Sun:2008nh}
Z.~Sun,
``Continuous degeneracy of non-supersymmetric vacua,''
Nucl. Phys. B \textbf{815} (2009), 240-255
[arXiv:0807.4000 [hep-th]].

\bibitem{Sun:2011aq}
Z.~Sun,
``Vacuum statistics and parameter tuning for F-term supersymmetry breaking,''
JHEP \textbf{09} (2011), 107
[arXiv:1105.3172 [hep-th]].

\bibitem{Shih:2007av}
D.~Shih,
``Spontaneous R-symmetry breaking in O'Raifeartaigh models,''
JHEP \textbf{02} (2008), 091
[arXiv:hep-th/0703196 [hep-th]].

\bibitem{Curtin:2012yu}
D.~Curtin, Z.~Komargodski, D.~Shih and Y.~Tsai,
``Spontaneous R-symmetry Breaking with Multiple Pseudomoduli,''
Phys. Rev. D \textbf{85} (2012), 125031
[arXiv:1202.5331 [hep-th]].

\bibitem{Azeyanagi:2012pc}
T.~Azeyanagi, T.~Kobayashi, A.~Ogasahara and K.~Yoshioka,
``Runaway, D term and R-symmetry Breaking,''
Phys. Rev. D \textbf{86} (2012), 095026
[arXiv:1208.0796 [hep-ph]].
  
\bibitem{Vaknin:2014fxa}
T.~Vaknin,
``New phases in O`Raifeartaigh-like models and R-symmetry breaking,''
JHEP \textbf{09} (2014), 004
[arXiv:1402.5851 [hep-th]].

\bibitem{Carpenter:2008wi}
L.~M.~Carpenter, M.~Dine, G.~Festuccia and J.~D.~Mason,
``Implementing General Gauge Mediation,''
Phys. Rev. D \textbf{79} (2009), 035002
[arXiv:0805.2944 [hep-ph]].

\bibitem{Sun:2008va}
Z.~Sun,
``Tree level spontaneous R-symmetry breaking in O'Raifeartaigh models,''
JHEP \textbf{01} (2009), 002
[arXiv:0810.0477 [hep-th]].

\bibitem{Liu:2014ida}
F.~Liu, M.~Liu and Z.~Sun,
``No-go for tree-level R-symmetry breaking,''
Eur. Phys. J. C \textbf{77} (2017) no.11, 745
[arXiv:1412.0183 [hep-ph]].

\bibitem{Cheung:2007es}
C.~Cheung, A.~L.~Fitzpatrick and D.~Shih,
``(Extra)ordinary gauge mediation,''
JHEP \textbf{07} (2008), 054
[arXiv:0710.3585 [hep-ph]].

\bibitem{Marques:2009yu}
D.~Marques,
``Generalized messenger sector for gauge mediation of supersymmetry breaking and the soft spectrum,''
JHEP \textbf{03} (2009), 038
[arXiv:0901.1326 [hep-ph]].

\bibitem{Giudice:1997ni}
G.~F.~Giudice and R.~Rattazzi,
``Extracting supersymmetry breaking effects from wave function renormalization,''
Nucl. Phys. B \textbf{511} (1998), 25-44
[arXiv:hep-ph/9706540 [hep-ph]].

\bibitem{Komargodski:2009jf}
Z.~Komargodski and D.~Shih,
``Notes on SUSY and R-Symmetry Breaking in Wess-Zumino Models,''
JHEP \textbf{04} (2009), 093
[arXiv:0902.0030 [hep-th]].

\bibitem{Abel:2009ze}
S.~A.~Abel, J.~Jaeckel and V.~V.~Khoze,
``Gaugino versus Sfermion Masses in Gauge Mediation,''
Phys. Lett. B \textbf{682} (2010), 441-445
[arXiv:0907.0658 [hep-ph]].

\bibitem{Intriligator:2006dd}
K.~A.~Intriligator, N.~Seiberg and D.~Shih,
``Dynamical SUSY breaking in meta-stable vacua,''
JHEP \textbf{04} (2006), 021
[arXiv:hep-th/0602239 [hep-th]].

\bibitem{Intriligator:2007py}
K.~A.~Intriligator, N.~Seiberg and D.~Shih,
``Supersymmetry breaking, R-symmetry breaking and metastable vacua,''
JHEP \textbf{07} (2007), 017
[arXiv:hep-th/0703281 [hep-th]].

\bibitem{Dreiner:2008tw}
H.~K.~Dreiner, H.~E.~Haber and S.~P.~Martin,
``Two-component spinor techniques and Feynman rules for quantum field theory and supersymmetry,''
Phys. Rept. \textbf{494} (2010), 1-196
[arXiv:0812.1594 [hep-ph]].

\bibitem{Peskin:1995ev}
M.~E.~Peskin and D.~V.~Schroeder,
``An Introduction to quantum field theory,''
Addison-Wesley, 1995,
ISBN 978-0-201-50397-5

\bibitem{Arkani-Hamed:2004ymt}
N.~Arkani-Hamed and S.~Dimopoulos,
``Supersymmetric unification without low energy supersymmetry and signatures for fine-tuning at the LHC,''
JHEP \textbf{06} (2005), 073
[arXiv:hep-th/0405159 [hep-th]].

\bibitem{Giudice:2004tc}
G.~F.~Giudice and A.~Romanino,
``Split supersymmetry,''
Nucl. Phys. B \textbf{699} (2004), 65-89
[erratum: Nucl. Phys. B \textbf{706} (2005), 487-487]
[arXiv:hep-ph/0406088 [hep-ph]].

\bibitem{Arkani-Hamed:2004zhs}
N.~Arkani-Hamed, S.~Dimopoulos, G.~F.~Giudice and A.~Romanino,
``Aspects of split supersymmetry,''
Nucl. Phys. B \textbf{709} (2005), 3-46
[arXiv:hep-ph/0409232 [hep-ph]].

\bibitem{Harigaya:2013asa}
K.~Harigaya, M.~Ibe and T.~T.~Yanagida,
``A Closer Look at Gaugino Masses in Pure Gravity Mediation Model/Minimal Split SUSY Model,''
JHEP \textbf{12} (2013), 016
[arXiv:1310.0643 [hep-ph]].

\bibitem{Wang:2013rba}
F.~Wang, W.~Wang and J.~M.~Yang,
``Split supersymmetry under GUT and current dark matter constraints,''
Eur. Phys. J. C \textbf{74} (2014) no.10, 3121
[arXiv:1310.1750 [hep-ph]].

\bibitem{Wang:2015mea}
F.~Wang, W.~Wang and J.~M.~Yang,
``A split SUSY model from SUSY GUT,''
JHEP \textbf{03} (2015), 050
[arXiv:1501.02906 [hep-ph]].

\bibitem{Ahmed:2019xon}
W.~Ahmed, A.~Mansha, T.~Li, S.~Raza, J.~Roy and F.~Z.~Xu,
``The upper bounds on supersymmetry breaking scales in split supersymmetry from the Higgs mass and electroweak vacuum stability,''
PTEP \textbf{2020} (2020) no.4, 043B08
[arXiv:1901.05278 [hep-ph]].

\bibitem{Cheung:2010mc}
C.~Cheung, Y.~Nomura and J.~Thaler,
``Goldstini,''
JHEP \textbf{03} (2010), 073
[arXiv:1002.1967 [hep-ph]].

\bibitem{Cheng:2010mw}
H.~C.~Cheng, W.~C.~Huang, I.~Low and A.~Menon,
``Goldstini as the decaying dark matter,''
JHEP \textbf{03} (2011), 019
[arXiv:1012.5300 [hep-ph]].

\bibitem{Cao:2020oxq}
J.~Cao, X.~Du, Z.~Li, F.~Wang and Y.~Zhang,
``Explaining The XENON1T Excess With Light Goldstini Dark Matter,''
[arXiv:2007.09981 [hep-ph]].

\end{thebibliography}
\end{document}